\documentclass[final,5p,times,twocolumn]{elsarticle}

\usepackage[colorlinks,allcolors={blue}]{hyperref}
\usepackage{amsmath}
\usepackage{natbib}
\usepackage{bm}
 
\usepackage{subfigure}
\usepackage{graphicx}
\usepackage{CJK}
\usepackage{xspace}  
\usepackage{multirow} 
\usepackage{dcolumn} 
\usepackage{pstricks}
\usepackage[utf8]{inputenc}
\usepackage{psfrag}
\usepackage[T1]{fontenc}
\usepackage{booktabs}
\usepackage{url}
\usepackage[normalem]{ulem}    

\usepackage[normalem]{ulem}

\journal{Physics Letters B}

\begin{document}

\begin{frontmatter}



\title{Identifying a characterized energy level structure of higher charmonium well matched to the peak structures in $e^+e^-\to \pi^+ D^0 D^{*-}$}

\author[pku,lzu1]{Jun-Zhang Wang}
\ead{wangjzh2022@pku.edu.cn}

\author[lzu1,lzu2,lzu3,lzu4]{Xiang Liu}
\ead{xiangliu@lzu.edu.cn}

\affiliation[pku]{organization={School of Physics and Center of High Energy Physics, Peking University},
            city={Beijing},
            postcode={100871}, 
            country={China}}
\affiliation[lzu1]{organization={School of Physical Science and Technology, Lanzhou University},
            city={Lanzhou},
            postcode={730000}, 
            country={China}}   
\affiliation[lzu2]{organization={Lanzhou Center for Theoretical Physics, Key Laboratory of Quantum Theory and Applications of MoE, Key Laboratory of Theoretical Physics of Gansu Province, Lanzhou University},
            city={Lanzhou},
            postcode={730000}, 
            country={China}}  
\affiliation[lzu3]{organization={Moe Frontiers Science Center for Rare Isotopes, Lanzhou University},
            city={Lanzhou},
            postcode={730000}, 
            country={China}}   
\affiliation[lzu4]{organization={Research Center for Hadron and CSR Physics, Lanzhou University $\&$ Institute of Modern Physics of CAS},
            city={Lanzhou},
            postcode={730000}, 
            country={China}}

\begin{abstract}

Recent progresses on charmoniumlike state have significantly enriched the discovery of new hadronic states, providing exciting opportunities for further investigations into the fascinating realm of charmonium physics. In this letter, we focus on the vector charmonium family and perform a detailed analysis of the recently observed $e^+e^-\to \pi^+ D^0 D^{*-}$ process. Our findings demonstrate a agreement between the observed peak structures and the predicted characterized energy level structure of higher vector charmonia including the $\psi(4220)$, $\psi(4380)$, $\psi(4415)$, and $\psi(4500)$, which are derived from an unquenched potential model. This discovery challenges conventional understanding of higher charmonia above 4 GeV and offers fresh insights into the dynamics of charm and anti-charm quarks in the formation of these states. Furthermore, the identification of these higher charmonia in the precisely measured $\pi^+ D^0D^{*-}$ open-charm decay channel would serve as compelling evidence supporting the unquenched scenario and contribute to a deeper understanding of the nonperturbative aspects of the strong interaction.
\end{abstract}




\end{frontmatter}




\section{Introduction}
\label{introduction}

As we enter the 21st century, higher precision experiments have yielded a plethora of surprising and promising results in hadron spectroscopy. Notably, a series of new hadronic states, such as the charmoniumlike $XYZ$ and $P_c$ states, have been discovered \cite{ParticleDataGroup:2020ssz,Chen:2016qju,Liu:2019zoy,Chen:2022asf,Guo:2017jvc,Olsen:2017bmm,Brambilla:2019esw}. This marks the onset of a new phase of hadron spectroscopy research, which may provide valuable insights into the nonperturbative problem of the strong interaction, a long-standing puzzle in particle physics.

In the 1970s, the observation of the $J/\psi$ particle \cite{E598:1974sol,SLAC-SP-017:1974ind} introduced a new class of hadrons known as charmonium. Subsequently, experimentalists discovered numerous charmonium states, which now constitute the bulk of charmonium cataloged by the Particle Data Group (PDG) \cite{ParticleDataGroup:2020ssz}. Moreover, these observations of charmonium stimulated the development of the Cornell model with the linear plus Coulomb potential, a quenched potential model proposed by Eichten {\it et al.} \cite{Eichten:1974af,Eichten:1978tg}. This model has inspired various potential models, including  famous Godfrey-Isgur (GI) model \cite{Godfrey:1985xj}, which facilitates quantitative calculations of hadron spectroscopy. { In addition to potential model approach, lattice QCD and effective field theory (such as potential NRQCD) have also made some progress on the excited charmonium spectroscopy, especially for predictions of charmonium hybrids \cite{Bali:2000gf,Ehmann:2007hj,Bali:2011rd,HadronSpectrum:2012gic,Cheung:2016bym,Lang:2015sba,Chen:2016ejo,Berwein:2015vca,Brambilla:1999xf}. }

Although substantial progress has been made in the field of hadron spectroscopy, it is important to note that our current understanding of it is not yet complete. In particular, establishing the energy level structure of higher charmonium states above 4.2 GeV has remained a challenge for several decades since the discovery of the $J/\psi$.
In 2014, the Lanzhou group drew attention to the similarity between the charmonium and bottomonium families \cite{He:2014xna}. They determined the mass of the narrow-width charmonium state $\psi(4S)$ to be 4263 MeV, which is markedly different from the conventional assignment of $\psi(4415)$ as the $\psi(4S)$ under quenched potential models such as the Cornell model \cite{Eichten:1974af,Eichten:1978tg}. However, this result is consistent with the expectations from the screening potential \cite{Dong:1994zj,Li:2009zu}.

Soon after, the BESIII Collaboration surprised us with the discovery of narrow structures around 4.2 GeV referred to as the $Y(4220)$ in processes involving $e^+e^-$ annihilations into hidden-charm channels, such as $J/\psi\pi^+\pi^-$ \cite{BESIII:2016bnd}, $h_c\pi^+\pi^-$ \cite{BESIII:2016adj}, $\omega\chi_{c0}$ \cite{BESIII:2019gjc}, and $\psi(3686)\pi^+\pi^-$ \cite{BESIII:2017tqk}. These observations were consistent with the expectation that there is a charmonium state, the $\psi(4220)$ \cite{Chen:2014sra}. Taking this opportunity, researchers realized the potential of the observed $Y(4220)$ narrow structure as a scaling point for constructing the energy level structure of higher charmonia above 4.2 GeV in Ref. \cite{Wang:2019mhs}. By using an unquenched potential model, the $Y(4220)$ was assigned to the charmonium family under the $4S$-$3D$ mixing scheme, and its charmonium partner $\psi(4380)$ was predicted \cite{Wang:2019mhs}. During this process, the well-established $\psi(4415)$ was proposed as a $5S$-$4D$ mixing state, while its charmonium partner $\psi(4500)$ was also found to exist \cite{Wang:2019mhs,Wang:2022jxj}. As a result, a characterized energy level structure for higher vector charmonia above 4.2 GeV has been obtained under an unquenched potential model. Given this research background, identifying this predicted higher charmonium energy level structure in its open-charm decay channels has become a critical step in the entire study. Open-charm decay channels are known to dominate the widths of higher charmonia, and their contributions may provide crucial evidence to test such a scenario.

In recent years, the BESIII Collaboration has made significant progress in measuring the $e^+e^-$ annihilation processes into open-charm channels \cite{BESIII:2018iea,BESIII:2021yvc,BESIII:2022quc,BESIII:2023cmv,BESIII:2023wsc}. In this letter, it is reported that the peak structures observed in the newly observed $e^+e^-\to \pi^+ D^0 D^{*-}$ process match well with the characterized energy level structure of higher vector charmonia predicted by the unquenched potential model, which includes the $\psi(4220)$, $\psi(4380)$, $\psi(4415)$, and $\psi(4500)$ states. This finding challenges our long-held perception of higher charmonia above 4 GeV and sheds new light on how charm and anti-charm quarks interact to form these states. The identification of these key higher charmonia in the $\pi^+ D^0 D^{*-}$ open-charm decay channel not only provides crucial evidence to test the unquenched scenario but also enriches our understanding of the nonperturbative aspects of the strong interaction.

\section{The emergence of the charmonium structures with characterized energy level in $e^+e^-\to \pi^+ D^0 D^{*-}$}

Building on the significance of the unquenched effect on the color confinement potential emphasized by the discovery of charmoniumlike $XYZ$ states \cite{Chen:2016qju}, we have directed our attention to the challenging charmoniumlike $Y$ problem by proposing a promising solution based on an unquenched quark potential model \cite{Wang:2019mhs}. As a result, our model predicted a comprehensive characterized energy spectrum for vector charmonia within the 4.2-4.6 GeV range \cite{Wang:2019mhs}, diverging from conventional quark model predictions \cite{Eichten:1974af,Eichten:1978tg}. To explore this characterized energy level structure, the open-charm reaction $e^+e^-\to \pi^+ D^0 D^{*-}$ should be a suitable process.  

The reaction dynamics of $e^+e^-\to \pi^+ D^0 D^{*-}$ is dominated by two-part contributions, i.e., the direct production and intermediate charmonium contribution. The amplitude of direct production can be described by a parameterized form 
\begin{eqnarray}
    \mathcal{A}_{0}=g\left(\sqrt{s}-m_D-m_{D^*}-m_{\pi}\right)^2e^{-a_0s}. \label{eq1}
\end{eqnarray}
And then the amplitude associated with intermediate charmonium can be written as  
\begin{eqnarray}
\mathcal{A}_{i}=\sqrt{12\pi R_i\Gamma_i}B(s,m_i,\Gamma_i)\sqrt{\Phi(\sqrt{s})/\Phi(m_i)}e^{i\phi_i},~ \label{eq:1}
\end{eqnarray}
where $B(s,m_i,\Gamma_i)=(s-m_i^2+im_i\Gamma_i)^{-1}$ is the Breit-Wigner function, and $m_i$, $\Gamma_i$, $\Phi(\sqrt{s})=\int32(\sqrt{s})^3/(2\pi)^3dm_{12}^2dm_{23}^2$ and $\phi_i$ are the mass and width of the intermediate charmonium state, three-body phase space and phase angle, respectively. The $R_i$ is the product of di-lepton width $\Gamma(\psi\to e^+e^-)$ and branching ratio $\mathcal{B}(\psi\to\pi^+ D^0 D^{*-})$. The corresponding cross section distribution against the center-of-mass energy $\sqrt{s}$ is 
\begin{eqnarray}   \sigma(s)=\left|\mathcal{A}_{0}(s)+\sum_i\mathcal{A}_{i}(s)\right|^2. \label{eq2}
\end{eqnarray}
{ Here, The $\mathcal{A}_0(s)+\sum_i\mathcal{A}_i(s)$ is total scattering amplitude of $e^+e^- \to D^0D^{*-}\pi^+$ after integrating the phase space, which should involve all possible partial-wave amplitudes in principle. It is worth mentioning that a recent lattice study indicates the importance of coupled-channel scattering in understanding scalar and tensor charmonium spectrum \cite{Wilson:2023hzu}, so we expect  that Eq. (\ref{eq2}) can be improved by a complete coupled-channel formalism in future studies. }
We noticed that the Born cross section of $e^+e^- \to\pi^+ D^0 D^{*-}$ between 4.05 and 4.60 GeV had been released by the BESIII Collaboration \cite{BESIII:2018iea}. In their analysis,  two broad enhancements were observed in the invariant mass spectrum of $\pi^+ D^0 D^{*-}$,  specifically around 4.2 and 4.4 GeV. However, the origin of these enhancements remains unclear at present. This presents us with an excellent opportunity to identify the predicted characterized energy spectrum of higher vector charmonium. 

{ Within the experimentally measured energy range, the unquenched potential model predicted four distinct underlying charmonium structures \cite{Wang:2019mhs}, in addition to the well-established charmonium state $\psi(4160)$. } The presence of multiple resonances and their interference can often lead to a serious multiple solutions problem in concrete analyses \cite{Yuan:2009gd,Zhu:2011ha,Bai:2019jrb}, which has to be carefully considered here. Starting from a general amplitude function in Eq. (\ref{eq2}), it can be written as
\begin{eqnarray}
\mathcal{A}(s)=\left(z_0D(s)+\sum_{k=1}^{k=n}z_kB_k(s)\right)\Phi(\sqrt{s})^{1/2},
\end{eqnarray}
where $D(s)$ is the direct amplitude function involving the energy-dependent term and $B_k(s)=B(s,m_k,\Gamma_k)=(s-p_k)^{-1}$ with $p_k=m_k^2-im_k\Gamma_k$. If the amplitude function $\mathcal{A}(s)$ has a series of zero points $\{x_m\}$ after making an analytical continuation of $s$ in the complex plane, by applying the Weierstrass factorization theorem \cite{Knopp}, it can be factorized as
\begin{eqnarray}
\mathcal{A}(s)=\prod_m(s-x_m)\mathcal{N}\left(s,\{z_0,z_k\}\right)\Phi(\sqrt{s})^{1/2}, \label{eq4}
\end{eqnarray}
where $\mathcal{N}$ function does not include any zero structures. Now our purpose is to search for a other set of solution $\{z_0^{\prime},z_k^{\prime}\}$, which satisfies 
\begin{eqnarray}
   \left|\prod_m(s-x_m)\mathcal{N}(s)\right|=\left|\prod_m(s-x_m^{\prime})\mathcal{N}^{\prime}(s)\right| \label{eq5}
\end{eqnarray}
for any $s$ in a real range. Due to the disparate structure in Eq. (\ref{eq5}), one can conclude that  $ |(s-x_m)|=|(s-x_m^{\prime})|$ and $|\mathcal{N}(s)|=|\mathcal{N}^{\prime}(s)|$. Here, it can be firstly known that $x_m^{\prime}=x_m$ or its complex conjugate $x_m^{*}$.
Additionally, by applying the identical relation 
\begin{eqnarray}
   && B_k(s)=-(s-x_m)B_k(s)/(x_m-p_k)+B_k(x_m), 
\end{eqnarray}
the $\mathcal{N}(s)$ in Eq. (\ref{eq4}) can be simplified as
\begin{eqnarray}
\mathcal{N}(s,\{z_0,z_k\})&=&\mathcal{C}(z_0,D(s),\{D(x_m)\}) \nonumber \\
&&+(-1)^{n}\sum_{k=1}^{n}\frac{z_k}{\prod_m(x_m-p_k)}B_k(s).
\end{eqnarray}
Because $\mathcal{N}$ function has no any zeros, in which each components are linearly independent, so the equation $|\mathcal{N}(s)|=|\mathcal{N}^{\prime}(s)|$ means that
\begin{eqnarray}
\left|\mathcal{C}(z_0,D(s),\{D(x_m)\})\right|&=&\left|\mathcal{C}(z_0^{\prime},D(s),\{D(x_m^{\prime})\})\right|, \label{eq8} \\
\left|\frac{z_k}{\prod_m(x_m-p_k)}\right|&=&\left|\frac{z_k^{\prime}}{\prod_m(x_m^{\prime}-p_k)}\right|. \label{eq9}
\end{eqnarray}
These two equations play a crucial role in analyzing the issue of multiple solutions caused by resonance interference. In Eq. (\ref{eq9}), if we consider the direct continuum contribution in $e^+e^- \to\pi^+ D^0 D^{*-}$ to be negligible and therefore ignored, we would anticipate obtaining 16 sets of $\{z_k^{\prime}\}$ solutions with identical fitting quality by making different replacements of $x_m^{\prime}$ with $x_m^*$. These solutions are nearly impossible to distinguish using various theoretical models. However, it is important to note that Eq. (\ref{eq8}) is generally not easily satisfied by the majority of the continuum term.  In the following, it can be seen that the direct continuum contribution in our fit is significant, enabling us to overcome the challenge of multiple solutions.

\begin{figure}[t]
\includegraphics[width=8.5cm,keepaspectratio]{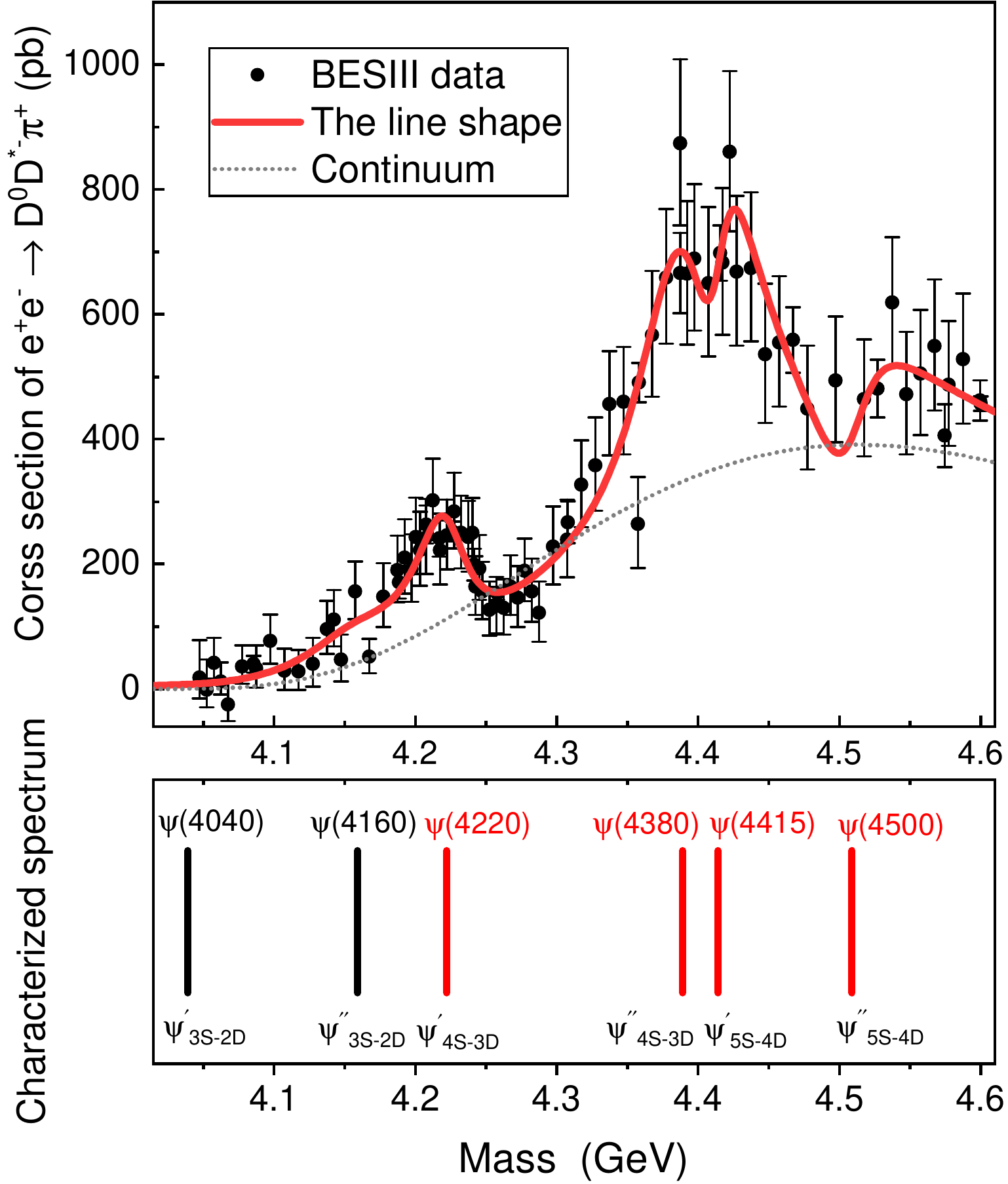}
	\caption{ The emergence of the characterized energy level structures of higher vector  charmonia in the cross section of $e^+e^- \to\pi^+ D^0 D^{*-}$. Here, the black lines and red lines in the bottom plot represent for the established charmonia and the predicted characterized charmonium structures in the unquenched scheme, respectively. \label{fig:Character}  }
\end{figure}

The fitted cross section distribution of $e^+e^- \to\pi^+ D^0 D^{*-}$ is depicted in Fig. \ref{fig:Character}, and the corresponding fitted parameters are summarized in Table. \ref{table:parameter}. It is evident that the experimental data can be well reproduced with clear signals of the four characterized energy level structures, as indicated by the fitted $\chi^2/d.o.f.=0.74$. Notably, the BESIII's study concluded that the second high-mass enhancement cannot be explained by a single resonance and possesses a more intricate underlying structure \cite{BESIII:2018iea}. In this regard, the characterized energy spectrum of charmonium provides a very natural explanation for this phenomenon. Specifically, the interference between the $\psi(4415)$ and the predicted characterized structure $\psi(4380)$ near 4.4 GeV can generate an extremely broad enhancement. Furthermore, BESIII has suggested categorizing the first enhancement as $Y(4220)$ \cite{BESIII:2018iea}, but its measured width of $77\pm6.8\pm6.3$ MeV \cite{BESIII:2018iea} obviously deviates the narrow width of $44.1\pm4.3\pm2.0$ $(28.2\pm3.9\pm3.6)$ MeV in hidden-charm channel $J/\psi\pi^+\pi^-(\chi_{c0}\omega)$ \cite{BESIII:2016bnd,BESIII:2019gjc} as well as the theoretical prediction of $(23\sim30)$ MeV \cite{Wang:2019mhs} for the charmonium $\psi(4220)$. However, our present analysis has convincingly proved that the first enhancement can be described within a multi-resonance interference scheme, with a narrow width of $\Gamma(\psi(4220))=44$ MeV as an input. As a result, this disagreement in widths can also be  effectively resolved.
From this perspective, we can confidently conclude that the enhancement structures observed in $e^+e^- \to\pi^+ D^0 D^{*-}$ indeed match well our prediction for the characterized energy spectrum of higher vector charmonia. This alignment serves as a strong indication for unraveling the nature of several reported charmoniumlike $Y$ states.   From Fig. \ref{fig:Character} and Table. \ref{table:parameter}, it can be also seen that the continuum contribution is highly significant and the only one set of solution for $\{R_i,\phi_i\}$ has been found when adopting the background form of $\mathcal{A}_0$ in Eq. (\ref{eq1}). This result implies that it will be convenient to continuously identify the characterized energy level structures of charmonia by comparing the fitted $R_i$ values with the corresponding theoretical three-body $\pi^+ D^0 D^{*-}$ decay behaviors, which will be discussed in detail later.

\section{The three-body $D D^{*}\pi$ decay behaviors of higher vector charmonia}

The fitted $R_i=\Gamma_{e^+e^-}^{\psi_i}\mathcal{B}(\psi_i\to\pi^+ D^0 D^{*-})$ directly quantifies the magnitude of the intermediate charmonium contribution. Thus, this physical quantity encodes important information for understanding the properties of higher charmonium states.  Very interestingly, one can find that the magnitudes of $R_i$ for $i=\psi(4040)$, $\psi(4220)$, $\psi(4415)$ and $\psi(4500)$ are approximately close to the range of $1.0\sim3.0$ eV as listed in Table \ref{table:parameter}, while the $R_{\psi(4380)}=19.0\pm4.6$ eV is significantly larger. For probing this phenomenon, we need to quantitatively calculate  the three-body $D D^{*}\pi$ decay behaviors of higher charmonia.

\subsection{The dynamical model for depicting the $e^+e^- \to \psi \to \pi^+D^0D^{*-}$ process}

The open-charm reaction $e^+e^- \to \psi \to \pi^+D^0D^{*-}$ is dominated by both the production of charmonium $\psi$ from the $e^+e^-$ annihilation and the open-charm decay of $\psi \to \pi^+D^0D^{*-}$.
The production amplitude of $e^+e^- \to \psi$ can be written as 
\begin{eqnarray}
\mathcal{M}(e^+e^- \to \psi)&=&\sqrt{2M_{\psi}}\int\frac{d^3k}{(2\pi)^3}\psi^*(\mathbf{k})\frac{1}{2m_c} \nonumber \\
&&\times \mathcal{M}(e^+e^- \to c(\mathbf{k})\bar{c}(-\mathbf{k})),
\end{eqnarray}
where $\mathcal{M}(e^+e^- \to c(\mathbf{k})\bar{c}(-\mathbf{k}))$ is a free scattering amplitude from the QED vertex, and wave function $\psi^*(\mathbf{k})$ represents the hadronization process of $c\bar{c}$ { from the virtual photon, whose distribution can be determined by solving the Schr\"{o}dinger equation in the potential model.} Since this free amplitude with same quark polarization is independent of the momentum, thus this integral for the orbital angular momentum between $c$ and $\bar{c}$ of $L=0$ is proportional to $\int d^3k \psi^*(\mathbf{k})$=$\psi^*(\mathbf{r})|_{\mathbf{r}=0}$. In our work, taking charmonium $\psi(4220)$ and its partner state $\psi(4380)$ as examples, their production from the $e^+e^-$ annihilation can be written as
\begin{eqnarray}
\sigma=\int\frac{(2\pi)^4|\mathcal{M}(e^+e^- \to \psi)|^2}{4p_1\sqrt{s}}d\Phi_1(p_1+p_2,K),
\end{eqnarray}
which can also directly relate to the di-lepton width of $\psi$ state. Due to the $S-D$ mixing effect is considered, the widths of $Y(4220)$ and $Y(4380)$ decaying into $e^+e^-$ are estimated by 
\begin{eqnarray}
&&\Gamma(Y(4220)\to e^+e^-)= \nonumber \\ &&~~~\frac{16\pi_c^2\alpha\mathcal{C}}{M^2_{Y(4220)}}\left|\cos\theta R_{4S}(0)+\frac{5\sqrt{2}}{M^2_{Y(4220)}}\sin\theta R^{\prime\prime}_{3D}(0)\right|^2, \\
&&\Gamma(Y(4380)\to e^+e^-)=  \nonumber \\ &&\frac{16\pi_c^2\alpha\mathcal{C}}{M^2_{Y(4380)}}\left|-\sin\theta R_{4S}(0)+\frac{5\sqrt{2}}{M^2_{Y(4380)}}\cos\theta R^{\prime\prime}_{3D}(0)\right|^2, 
\end{eqnarray}
where $e_c=2/3$, $\alpha=1/137$ and $\mathcal{C}=(1-\frac{16\alpha_s}{3\pi})$ is the first-order QCD radiative correction factor. In section II of the main text, the production of charmonium is determined by $R_i=\Gamma(\psi \to e^+e^-)\times\mathcal{B}(\psi \to D^0D^{*-}\pi^+)$ in Eq. (\ref{eq:1}), whose multiple solutions are determined by the fit to the experimental cross section.  It can be seen that the coupling dynamics between $\psi$ and $e^+e^-$ absolutely depends on the $S$-wave radial wave function and the second derivative of the $D$-wave radial wave function  at the origin. In Ref. \cite{Wang:2019mhs}, the wave functions of $\psi(4220)$ and $\psi(4380)$ had been calculated, by which $\Gamma(Y(4220)\to e^+e^-)$=290 keV and $\Gamma(Y(4220)\to e^+e^-)$=260 keV were obtained, whose orders of magnitude are quite typical for the production of excited charmonium states from the $e^+e^-$ annihilation. 

The decay of $\psi \to D \bar{D}^{*}\pi+c.c. $ is generally achieved via the direct coupling and two kinds of cascade decays from intermediate charmed meson states, i.e., $\psi \to D \bar{D}^{***} +c.c. \to D (\bar{D}^{***}\to \bar{D}^{*}\pi) +c.c. $ and $\psi \to D^* \bar{D}^{**} +c.c.\to D^* (\bar{D}^{**}\to \bar{D}\pi)+c.c. $, where $\bar{D}^{***}$ or $\bar{D}^{**}$ is the allowed intermediate charmed meson state. Due to the absence of experimental information, it is hard to quantitatively discuss the direct coupling contribution. However, the two-body cascade decay contribution may be enough for us to make a rough discussion on the decay behaviors of $\pi^+ D^0 D^{*-}$ because the summation for theoretical two-body OZI-allowed strong decay widths of many heavy quarkonia are already consistent with the corresponding experimental total widths \cite{Barnes:2005pb,Godfrey:2015dia,Wang:2018rjg}. By consulting the two-body open charm decay behaviors of the characterized energy level structures of higher vector charmonia \cite{Wang:2019mhs} and the charmed meson family information \cite{Song:2015fha}, we mainly consider four intermediate charmed meson states, i.e., $D^{***}=D_1(2420), D_1(2430), D_2^*(2460)$ and $D^{**}=D_0^*(2300)$.

Within the effective Lagrangian approach \cite{Casalbuoni:1996pg}, the decay amplitudes of $\psi \to D^*D_0^*(2300)$, $\psi \to DD_1(2420)$, $\psi \to DD_1(2430)$ and $\psi \to DD_2^*(2460)$   are written as 
\begin{eqnarray}
   \mathcal{M}_{D_0}&=&\frac{g_{\psi D^*D_0}\epsilon_{\psi\mu}\epsilon_{D^*}^{*\mu}}{q^{\prime2}-m_{D_0}^2+im_{D_0}\Gamma_{D_0}}g_{D_0 D\pi},  \label{eq:15}\\
   \mathcal{M}_{D_1}&=&g_{\psi DD_1}\epsilon_{\psi\mu}\frac{-g^{\mu\nu}+q^{\mu}q^{\nu}/m_{D_1}^2}{q^2-m_{D_1}^2+im_{D_1}\Gamma_{D_1}}g_{D_1 D^*\pi}\epsilon_{D^*\nu}^*,  \\
    \mathcal{M}_{D_1^{\prime}}&=&g_{\psi DD_1^{\prime}}\epsilon_{\psi\mu}\frac{-g^{\mu\nu}+q^{\mu}q^{\nu}/m_{D_1^{\prime}}^2}{q^2-m_{D_1^{\prime}}^2+im_{D_1^{\prime}}\Gamma_{D_1^{\prime}}}g_{D_1^{\prime} D^*\pi}\epsilon_{D^*\nu}^*,  \\
  \mathcal{M}_{D_2}&=&g_{\psi DD_2}\varepsilon_{\rho\lambda\mu\nu}p^{\mu}(p-q_2)^{\nu}\epsilon_{\psi}^{\rho}(p+q_2)^{\alpha}g_{D_2 D^*\pi} \nonumber \\ && \times \frac{\mathcal{G}_{\alpha\beta}^{\lambda\tau}(q^{\prime2})\varepsilon_{\kappa\tau\gamma\chi}q_3^{\gamma}(p-q_2)^{\chi}\epsilon_{D^*}^{*\kappa}(q_4-q_3)^{\beta}}{q^{\prime2}-m_{D_2}^2+im_{D_2}\Gamma_{D_2}},  \label{eq:18}
\end{eqnarray}
respectively, where $p,q_2,q_3,$ and $q_4$ are four-momentum of $\psi, D, D^*, $ and $ \pi$, respectively, and $q^{\prime}=p-q_3$ and $ q=p-q_2$. The spin projection operator of tensor particle $\mathcal{G}_{\alpha\beta}^{\lambda\tau}(q^{\prime2})=\frac{1}{2}(\Tilde{g}_{\alpha}^{\tau}\Tilde{g}_{\beta}^{\lambda}+\Tilde{g}_{\alpha\beta}\Tilde{g}^{\tau\lambda})-\frac{1}{3}\Tilde{g}_{\alpha}^{\lambda}\Tilde{g}_{\beta}^{\tau}$, where $\Tilde{g}_{\alpha\beta}=-g_{\alpha\beta}+q^{\prime\alpha}q^{\prime\beta}/q^{\prime2}$. The coupling constant $g_{D_0 D\pi}=7.17$ GeV, $g_{D_1(2420) D^*\pi}=1.66$ GeV, $g_{D_1(2430) D^*\pi}=5.39$ GeV and $g_{D_2 D^*\pi}=5.11$ GeV$^{-2}$  can be obtained by the corresponding experimental width of $\Gamma(D_0^*(2300)\to D\pi)\approx229\pm16$ MeV \cite{ParticleDataGroup:2020ssz}, $\Gamma(D_1(2420)\to D^*\pi)\approx31.3\pm1.9$ MeV \cite{ParticleDataGroup:2020ssz},  $\Gamma(D_1(2430)\to D^*\pi)\approx314\pm29$ MeV \cite{ParticleDataGroup:2020ssz} and  $\Gamma(D_2^*(2460)\to D^*\pi)\approx(47.3\pm0.8)/2.52$ MeV \cite{ParticleDataGroup:2020ssz}, respectively.

Hence, the total decay amplitude of $\psi \to D^0D^{*-}\pi^{+}$ reads 
\begin{eqnarray}
\mathcal{M}&=&\mathcal{M}_{D_0}+e^{i\theta_1}\mathcal{M}_{D_1(2420)}+e^{i\theta_2}\mathcal{M}_{D_1(2430)} \nonumber\\
&&+e^{i\theta_3}\mathcal{M}_{D_2(2460),}
\end{eqnarray}
where $\theta_1$, $\theta_2$, $\theta_3$ are three unknown phase angles. 

\subsection{Results and discussion}

\begin{figure}[h]
\includegraphics[width=8.7cm,keepaspectratio]{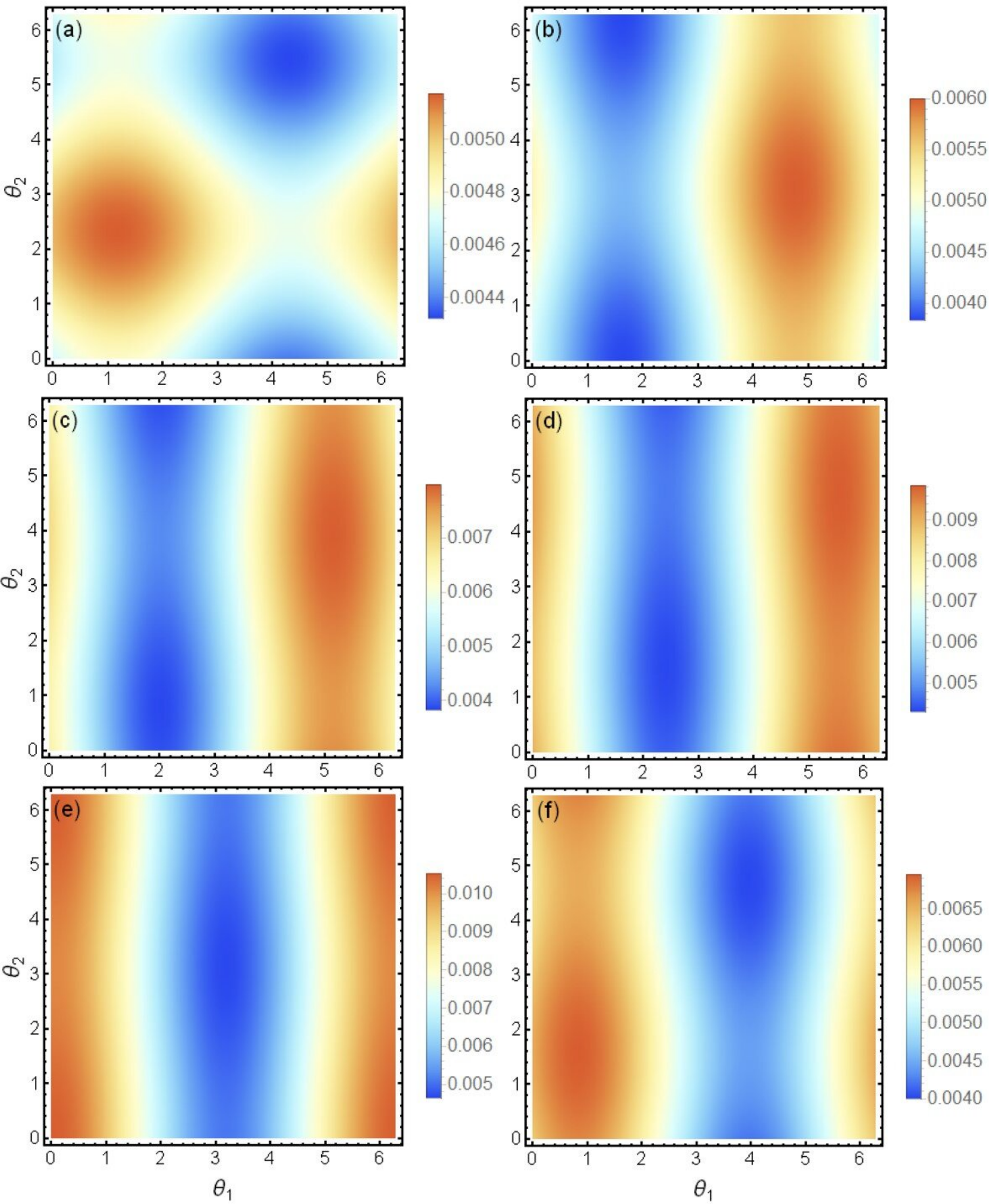}
	\caption{ The density plot of three-body decay width of $\psi(4380) \to \pi^{+} D^0D^{*-}$ as a function of the $\theta_1$, $\theta_2$ and $\theta_3$. Figure $(a)$ to $(f)$ corresponds to $\theta_3=0$, $\pi/4$, $(2\pi)/4$, $(3\pi)/4$, $(5\pi)/4$ and $(7\pi)/4$, respectively. The width value is in unit of GeV.  \label{fig:threebodydecay}  }
\end{figure}

The three-body $\pi^{+}D^0D^{*-}$ decay width can be calculated by 
\begin{eqnarray}
\Gamma_{3-body}=\int\frac{|\mathcal{M}|^2}{(2\pi)^516m_{\psi}^2}|q_2||q_3^*|d\Omega_2d\Omega_3^*dm_{D^*\pi},
\end{eqnarray}
where $q_3^*$ and $\Omega_3^*$ are the three momentum and solid angle of $D^{*-}$ in the center of the $D^{*-}\pi^{+}$ system mass frame, respectively. Based on the above formula, we firstly estimate the three-body decay width of $\psi(4380) \to \pi^{+}D^0D^{*-}$. The four related coupling constants $g_{\psi DD_0}=1.29$ GeV, $g_{\psi DD_1(2420)}=2.09$ GeV, $g_{\psi DD_1(2430)}=1.34$ GeV and  $g_{\psi DD_2}=1.97$ GeV$^{-2}$  can be fixed by the corresponding theoretically estimated two-body widths listed in Ref. \cite{Wang:2019mhs}. The width distribution plot of $\psi(4380) \to \pi^{+} D^0D^{*-}$ over the full range $[0\sim2\pi]$ of $\theta_1$, $\theta_2$ and $\theta_3$ is illustrated in Fig. \ref{fig:threebodydecay}. It can be seen that the three-body width is weakly dependent on $\theta_2$ and its minimum 3.7 MeV appears near $\theta_1=\frac{\pi}{2}\sim\frac{3\pi}{4}$ and $\theta_3=\frac{\pi}{4}\sim\frac{\pi}{2}$ and maximum 10.9 MeV occurs around $\theta_1=0$ and $\theta_3=\pi\sim\frac{5\pi}{4}$. Considering the di-lepton width of $\psi(4380)\to e^+e^-=260$ eV \cite{Wang:2019mhs}, we can estimate  $R_{\psi(4380)}=12.0\sim35.8$ eV, which aligns with the fitted value of $19.0\pm4.6$ eV listed in Table \ref{table:parameter}. Similarly, we calculated $\Gamma(\psi(4415)\to \pi^{+}D^0D^{*-})=(0.35\sim1.9)$ MeV and $\Gamma(\psi(4500)\to \pi^{+}D^0D^{*-})=(1.5\sim5.5)$ MeV, which can be converted to be $R_{\psi(4415)}=(2.4\sim13.5)$ eV and $R_{\psi(4500)}=(3.3\sim12.3)$ eV, respectively, using the input values of $\Gamma(\psi(4415)\to e^+e^-)=230$ eV \cite{BES:2007zwq,Wang:2019mhs} and $\Gamma(\psi(4500)\to e^+e^-)=113$ eV \cite{Wang:2019mhs}. { In fact, when describing the broad resonance such as $D_0^*$ and $D_1^{\prime}$, our formalism of Eqs. (\ref{eq:15})-(\ref{eq:18}) can be further improved by including the invariant mass-dependent width effect in the propagator. Concretely, the constant width $\Gamma_0$ in the propagator $(q^2-m^2+im\Gamma_0)^{-1}$ can be replaced to 
\begin{eqnarray}
\Gamma(q^2)=\Gamma_0\frac{m^2}{q^2}\left(\frac{\lambda(q^2,m_{D^{(*)}}^2,m_{\pi}^2)/q^2}{\lambda(m^2,m_{D^{(*)}}^2,m_{\pi}^2)/m^2}\right)^{(2l+1)/2},
\end{eqnarray}  
where $\lambda(x,y,z)=x^2+y^2+z^2-2xy-2xz-2yz$ is the K\"{a}llen function and $l$ is the orbital angular momentum for the intermediate excited charmed meson. Based on this effect, we recalculate the three-body decays of $\pi^+D^0D^{*-}$ for $\psi(4380)$, $\psi(4415)$ and $\psi(4500)$, which are summarized in Table \ref{tab:com}. It can be found that the results within using the invariant mass-dependent width are similar to those within using the constant width. The main reason for this consistency lies in the non-dominant contributions of the broad resonances $D_0^*$ and $D_1^{\prime}$.}

Remarkably, our theoretical results roughly confirm the dominant contribution of $\psi(4380)$ in the $e^+e^- \to\pi^+ D^0 D^{*-}$ process through self-consistency, lending strong support to the proposed characterized energy level structure of higher vector charmonia. Moreover, it is worth noting that the theoretically predicted signal of $\psi(4500)\to \pi^{+}D^0D^{*-}$ should exhibit a more pronounced line shape than the present data, which poses an intriguing question for future precise experimental investigations.
{ Finally, we would like to emphasize again the crucial role of the characterized energy level spectrum of charmonia in deciphering the cross section of $e^+e^- \to \pi^+D^0D^{*-}$. When excluding the $\psi(4380)$ from the fit, the previous analysis by the BESIII Collaboration has discovered a significantly large width $191.9\pm13.0$ MeV for the second high-mass enhancement, which is in stark contradiction with the width of $\psi(4415)$   \cite{BESIII:2018iea}. Similarly, when removing the contribution of $\psi(4160)$, the width $77\pm6.8\pm6.3$ MeV of the first enhancement in the cross section of $e^+e^- \to \pi^+D^0D^{*-}$ \cite{BESIII:2018iea} also obviously deviates the narrow width of $44.1\pm4.3\pm2.0$ $(28.2\pm3.9\pm3.6)$ MeV in hidden-charm channel $J/\psi\pi^+\pi^-(\chi_{c0}\omega)$ \cite{BESIII:2016bnd,BESIII:2019gjc} for the $Y(4220)$. Thus, these disagreement in widths can be  effectively resolved in our fitting scheme involving all characterized energy level structures of higher charmonium.}

\begin{table*}[t]
  	\caption{The fitted parameters of describing the cross section distribution of $e^+e^- \to\pi^+ D^0 D^{*-}$ based on the scheme involving the characterized energy level structures of higher charmonia, where the $\chi^2/d.o.f.=0.74$.}
  	\setlength{\tabcolsep}{4.0mm}{
  	\begin{tabular}{lccccccccccccc}
			\toprule[1.0pt]
   \toprule[1.0pt]
     Parameters & $a_0$ (GeV$^{-2}$)  & $g$ (GeV$^{-3}$) & $R_{\psi(4160)}$ (eV)  & $R_{\psi(4220)}$ (eV)  & $R_{\psi(4380)}$ (eV)  & $R_{\psi(4415)}$ (eV) 
     \\
		\toprule[1.0pt]
         Values  & $0.445\pm0.025$  & $34.5\pm18.1$  & $0.726\pm0.527$ & $2.70\pm0.63$ & $19.0 \pm4.6$ & $2.34\pm1.23$\\ 
         \toprule[1.0pt]
         Parameters &  $R_{\psi(4500)}$ (eV)  & $\phi_{\psi(4160)}$ (rad) &
     $\phi_{\psi(4220)}$ (rad) &
     $\phi_{\psi(4380)}$ (rad)&
     $\phi_{\psi(4415)}$ (rad)&
     $\phi_{\psi(4500)}$ (rad) \\
                         
                     \bottomrule[0.6pt]     
                     Values & $1.60\pm0.42$  & $1.97\pm0.65$ & $2.07\pm0.15$ & $1.44\pm0.16$ & $6.04\pm0.24$ &$5.76\pm0.31$  \\  
			\bottomrule[1.0pt]
           \bottomrule[1.0pt]
		\end{tabular}\label{table:parameter}}
  \end{table*}

\begin{table}[h]
\caption{The comparison of the three-body decay width using the constant width $\Gamma_0$ and invariant mass-dependent width $\Gamma(q^2)$ for the intermediate excited charmed meson. The results are in units of MeV.  \label{tab:com}}
	\setlength{\tabcolsep}{3.7mm}{
 \begin{tabular}{ccccc}
	\toprule[1pt]
 \toprule[1pt]
	Decay & $\Gamma_0$  & $\Gamma(q^2)$ \\
	\midrule[1pt]
 $\psi(4380)\to \pi^+D^0D^{*-}$ & $3.37\sim10.3$ & $3.62\sim11.4$  \\
 $\psi(4415)\to \pi^+D^0D^{*-}$ & $0.347\sim1.94$ &  $0.334\sim2.05$  \\
 $\psi(4500)\to \pi^+D^0D^{*-}$ & $1.48\sim5.47$ & $1.43\sim5.60$  \\
	\bottomrule[1pt]
 \bottomrule[1pt]
	\end{tabular}}
\end{table}

\section{A remark on the significance of discovering the characterized energy level structure of higher charmonia}

The characterized energy spectrum is a powerful way to distinguish different theoretical models in hadron physics. The most known example is the observation of the hidden-charm pentaquarks $P_c(4312)^+$, $P_c(4440)^+$ and $P_c(4457)^+$ by LHCb \cite{LHCb:2019kea}. Specifically, these three structures that all slightly below the threshold composed of a charmed baryon $\Sigma_c$ and an anti-charmed meson $\bar{D}(\bar{D}^{*}) $ can perfectly meet the characterized energy spectrum of  hidden-charm molecule of $S$-wave $\bar{D}^{(*)}\Sigma_c$, i.e., $\bar{D}\Sigma_c$ with $J^P=1/2^-$, $\bar{D}^*\Sigma_c$ with $J^P=1/2^-$ and $J^P=3/2^-$ \cite{Wu:2010jy,Wang:2011rga,Yang:2011wz,Uchino:2015uha,Karliner:2015ina,Chen:2022asf}.  As a result, the hadronic molecule explanation for $P_c$ states has been widely recognized \cite{Chen:2022asf}.

Return to the concerned charmoniumlike $Y$ states, at present, the most popular interpretations of exotic states for vector charmoniumlike states include hybrid charmonium and $S$-wave hidden-charm hadronic molecules \cite{Chen:2016qju}, whose spectrum behaviors are completely different in the narrow energy region between 4.2 and 4.5 GeV. Obviously, the characterized energy spectrum of hidden-charm hadronic molecules should be straightforward, which must be related to the threshold. 
There are five S-wave hidden-charm systems $DD_1(2420)$, $D^*D_1(2420)$, $D^*D_2(2460)$, $D_sD_{s1}(2460)$ and $D_s^*D_{s0}^{*}(2317)$ between 4.2 and 4.5 GeV. However, the $DD^*\pi$ channel is forbidden in the $D^*D_1(2420)$ system and also strongly suppressed in the hidden-charm system involving strange quark due to the OZI rule. Additionally, it is worth noting that the thresholds for $D^*D_2(2460)$, $D_sD_{s1}(2460)$ and $D_s^*D_{s0}^{*}(2317)$  are all approximately at 4.44 GeV. Consequently, one can anticipate that the characterized energy spectrum of hadronic molecule assumption would differ obviously from our framework.
Regarding the mass spectrum of hybrid charmonium $c\bar{c}g$ with the quantum number $J^{PC}=1^{--}$, lattice QCD calculations and several theoretical models support $Y(4220)$ as a hybrid candidate \cite{HadronSpectrum:2012gic,Zhu:2005hp,Close:2005iz,Kou:2005gt,Guo:2008yz,Segovia:2014mca}. 
Here, it is worth mentioning that the calculations of QCD sum rule for the $P$-wave hidden-charm tetraquark $cq\bar{c}\bar{q}$ and $cs\bar{c}\bar{s}$ with $J^{PC}=1^{--}$ shows a spectrum desert within the energy region between 4.2 and 4.5 GeV \cite{Chen:2010ze}.

Given the clear distinctiveness of the characterized energy level spectrum resulting from various internal configurations, it is undeniable that the discovery  of these characterized energy level structures for higher charmonia in this work offers very important insights into the nature of charmoniumlike $Y$ particles.

\section{Summary}

This letter aims to address the long-standing challenge of constructing the higher vector charmonium family above 4 GeV. By analyzing the open-charm process $e^+e^-\to \pi^+ D^0 D^{*-}$ in detail, we reveal a remarkable agreement between the observed enhancement structures and the characterized energy level structure predicted by the unquenched potential model. The identified states, including $\psi(4220)$, $\psi(4380)$, $\psi(4415)$, and $\psi(4500)$, challenge conventional understanding and offer new insights into the dynamics of charm and anti-charm quarks in their formation.  By overcoming multiple solution problems arising from resonance interference in the cross section analysis, we further facilitate identification of characterized energy level structures of higher vector charmonia by comparing them with theoretical predictions of the three-body  $\pi^+ D^0 D^{*-}$ decay behaviors. By utilizing the effective decay amplitude, the calculated three-body decay behaviors of $\psi \to \pi^+ D^0 D^{*-}$ sheds light on the properties of higher vector charmonia, particularly confirming the significant contribution of $\psi(4380)$ in $e^+e^- \to\pi^+ D^0 D^{*-}$. Collectively, these findings provide strong evidence for the existence of the characterized energy level structure of higher vector charmonia. This work deepens our understanding of charmonium physics, emphasizing the need for precise experimental measurements of $e^+e^- \to\pi^+ D^0 D^{*-}$ to further advance the study of charmonium spectroscopy above 4.0 GeV.


\section*{ACKNOWLEDGEMENTS}

This work is supported by the National Natural Science Foundation of China under Grant Nos. 12335001, 12247101 and 12047501, the China National Funds for Distinguished Young Scientists under Grant No. 11825503, the National Key Research and Development Program of China under Contract No. 2020YFA0406400, the 111 Project under Grant No. B20063, the Fundamental Research Funds for the Central Universities under Grant No. lzujbky-2023-stlt01, and the project for top-notch innovative talents of Gansu province. J.Z.W. is also supported by the National Postdoctoral Program for Innovative Talent.

\end{document}